\documentclass[12pt]{iopart}

\usepackage{lineno,hyperref}
\usepackage{iopams}
\usepackage{graphicx}  
\begin{document}

\title[Spin-Orbital momentum decomposition]{Spin-Orbital momentum decomposition and helicity exchange in a set of non-null knotted electromagnetic fields}

\author{M Array\'as and J L Trueba}

\address{\'Area de Electrogmanetismo, Universidad Rey Juan Carlos, Tulip\'an s/n, 28933 M\'ostoles, Madrid, Spain}
\ead{manuel.arrayas@urjc.es, joseluis.trueba@urjc.es}
\vspace{10pt}
\begin{indented}
\item[]June 26th 2017
\end{indented}

\begin{abstract}
We calculate analytically the spin-orbital decomposition of the angular momentum using completely non-paraxial fields that have certain degree of linkage of electric and magnetic lines. The split of the angular momentum into spin-orbital components is worked out for non-null knotted electromagnetic fields. The relation between magnetic and electric helicities and spin-orbital decomposition of the angular momentum is considered. We demonstrate that even if the total angular momentum and the values of the spin and orbital momentum are the same, the behaviour of the local angular momentum density is rather different. By taking cases with constant and non-constant electric and magnetic helicities, we show that the total angular momentum density present different characteristics during time evolution.
\end{abstract}


\section{Introduction}
There has been recently some interest in the orbital-spin decomposition of the angular momentum carried by light. The total angular momentum can be decomposed into orbital and spin angular momenta for paraxial light, but for nonparaxial fields that splitting is more controversial because their quantized forms do not satisfied the commutation relations \cite{vanEnk1,vanEnk2}. For review and references see for example \cite{Allen2003,Bliokh2012}. 

In this work, we provide an exact calculation of the orbital-spin decomposition of the angular momentum in a completely nonparaxial field. We compute the orbital-spin contributions to the total angular momentum analytically for a knotted class of fields \cite {Arr15}. These fields have nontrivial electromagnetic helicity \cite{Afana,True96}. We show that it is possible the existence of electromagnetic fields in vacuum with the same constant angular momentum and orbital-spin decomposition but different electric and magnetic helicities. We find cases where the helicities are constant during the field evolution, and cases where they change in time, evolving through a phenomena of exchanging magnetic and electric components \cite{Arr11}. The angular momentum density presents different time evolution in each case. 

The orbital-spin decomposition and its observability has been discussed in the context of the dual symmetry of Maxwell equations in vacuum \cite{Bliokh2013}. In this paper we first make a brief review of the concept of electromagnetic duality. That duality, termed ``electromagnetic democracy'' \cite{Berry}, has been central in the work of knotted field configurations \cite{Ran89,Ran90,Ran92,Ran95,Ran97,Ran98,Ran01,Irv08,Besieris,Arr10,Enk,Kedia,Arr15,CarlosHoyo,Kedia17,Arr17,Arr17b}. Related field configurations have also appeared in plasma physics \cite{Kamchatnov,Semenov2002,Bouw2014,Bouw2015}, optics \cite{Dennis09,Dennis10,Dennis11,Dennis12,Dennis17}, classical field theory \cite{Faddeev1997}, quantum physics \cite{Hall2016,Dennis16}, various states of matter \cite{Volovik1977,Dzyloshinskii,Kawaguchi2008,Irv2013,Irv2014} and twistors \cite{Dalh2012,Bouw2015b}.

We will make use of the helicity basis \cite{True96} in order to write the magnetic and electric spin of the field in that basis, which simplifies the calculations, as well as the magnetic and electric helicities components. In that basis we will get some general results such as the different between the magnetic and electric spin components in the Coulomb gauge is null. This conclusion coincides with the results found, for example, in \cite{Barnett2011} using a different approach. We proceed by giving the explicit calculation of the decomposition of the angular momentum in spin and orbital components for a whole class of fields, the non-null toroidal class \cite{Arr15,Arr17}. We will show that the angular decomposition remains constant in time, while the helicities may or may not change. We provide an example of each case and plot the time evolution of the total angular momentum density. In the final section we summarise the main results.

\section{Duality and helicity in Maxwell theory in vacuum}
In this section we will review the definition of magnetic and electric helicities. These definitions are possible because the dual property of electromagnetism in vacuum. We will also describe a vector density which can be identify with the spin density using the helicity four-currents zeroth component.

Electromagnetism in vacuum can be described in terms of two real vector fields, ${\bf E}$ and ${\bf B}$, called the electric and magnetic fields respectively. Using the SI of Units, these fields satisfy Maxwell equations in vacuum,
\begin{eqnarray}
\nabla \cdot {\bf B} =0&,& \, \, \nabla \times {\bf E} + \frac{\partial
{\bf B}}{\partial t} =0, \label{elmaghel1} \\
\nabla \cdot {\bf E} =0&,& \, \, \nabla \times {\bf B} - \frac{1}{c^2} \frac{\partial
{\bf E}}{\partial t} =0. \label{elmaghel2}
\end{eqnarray}
Using the four-vector electromagnetic potential
\begin{equation}
A^{\mu} = \left( \frac{V}{c} , {\bf A} \right),
\label{elmaghelpot} 
\end{equation}
where $V$ and ${\bf A}$ are the scalar and vector potential respectively, the electromagnetic field tensor is
\begin{equation}
F_{\mu \nu} = \partial_{\mu} A_{\nu} - \partial_{\nu} A_{\mu} .
\label{elmaghelfield}
\end{equation}
From (\ref{elmaghelfield}), the electric and magnetic field components are
\begin{equation}
{\bf E}_{i} = c \, F^{i0} , \, \, {\bf B}_{i} = - \frac{1}{2} \epsilon_{ijk} F^{jk} ,
\label{elmaghelfield1}
\end{equation}
or, in three-dimensional quantities,
\begin{equation}
{\bf E} = - \nabla V - \frac{\partial {\bf A}}{\partial t}, \, \, \,  {\bf B} = \nabla \times {\bf A} .
\label{elmaghelpot1}
\end{equation}
Since equations (\ref{elmaghel1}) are just identities in terms of the four-vector electromagnetic potential (\ref{elmaghelpot}), by using (\ref{elmaghelpot1}), the dynamics of electromagnetism is given by equations (\ref{elmaghel2}), that can be written as
\begin{equation}
\partial _{\mu} F^{\mu \nu} = 0 .
\label{elmaghel4}
\end{equation}
Partly based on the duality property of Maxwell equations in vacuum \cite{Stratton} there is the idea of ``electromagnetic democracy'' \cite{Berry,Bliokh2013}. The equations are invariant under the map $({\bf E} , c{\bf B} ) \mapsto (c{\bf B} , -{\bf E} )$. Electromagnetic democracy means that, in vacuum, it is possible to define another four-potential
\begin{equation}
C^{\mu} = (c \, V^{\prime}, {\bf C}),
\label{cdefinition}
\end{equation}
so that the dual of the electromagnetic tensor $F_{\mu \nu}$ in equation (\ref{elmaghelfield}), defined as
\begin{equation}
{}^{*}\!F_{\mu \nu} = \frac{1}{2} \varepsilon_{\mu \nu \alpha \beta} F^{\alpha \beta} ,
\label{elmaghel5}
\end{equation}
satisfies
\begin{equation}
{}^{*}\!F_{\mu \nu} =-\frac{1}{c} \left( \partial_{\mu} C_{\nu} - \partial_{\nu} C_{\mu} \right) ,
\label{elmaghel7}
\end{equation}
or, in terms of three-dimensional fields,
\begin{equation}
{\bf E} = \nabla \times {\bf C} , \, \, \, {\bf B} = \nabla V^{\prime} + \frac{1}{c^2} \frac{\partial {\bf C}}{\partial t} .
\label{elmaghel8}
\end{equation}
Equations (\ref{elmaghel2}) are again identities when definitions (\ref{elmaghel8}) are imposed. Thus Maxwell equations in vacuum can be described in terms of two sets of vector potentials as in definitions (\ref{elmaghelfield}) and (\ref{elmaghel7}), that have to satisfy the duality condition (\ref{elmaghel5}). 

In the study of topological configurations of electric and magnetic lines, an important quantity is the helicity of a vector field \cite{Moffatt,Berger,Ricca,Berger1999,Dennis05,Ricca2011}, that can be defined for every divergenceless three-dimensional vector field. Magnetic helicity is related to the linkage of magnetic lines. In the case of electromagnetism in vacuum, the magnetic helicity can be defined as the integral
\begin{equation}
h_{m}= \frac{1}{2 c \mu_{0}} \int d^3 r \, {\bf A} \cdot {\bf B} ,
\label{elmaghel10}
\end{equation}
where $c$ is the speed of light in vacuum and $\mu_{0}$ is the vacuum permeability. Note that, in this equation, the magnetic helicity is taken so that it has dimensions of angular momentum in the SI of Units. Since the electric field in vacuum is also divergenceless, an electric helicity, related to the linking number of electric lines, can also be defined as
\begin{equation}
h_{e} = \frac{\varepsilon_{0}}{2c} \int d^3 r \, {\bf C} \cdot {\bf E} = \frac{1}{2c^3 \mu_{0}} \int d^3 r \, {\bf C} \cdot {\bf E},
\label{elmaghel11}
\end{equation}
where $\varepsilon_{0} = 1/(c^2 \mu_{0} )$ is the vacuum electric permittivity. Electric helicity in equation (\ref{elmaghel11}) also has dimensions of angular momentum. Magnetic and electric helicities in vacuum can be studied in terms of helicity four-currents \cite{Afana,True96,Ran01,Bliokh2013}, so that the magnetic helicity density is the zeroth component of
\begin{equation}
{\cal H}_{m}^{\mu} =  -\frac{1}{2 c \mu_{0}} \, A_{\nu} {}^{*}\!F^{\nu \mu},
\label{elmaghel12}
\end{equation}
and the electric helicity is the zeroth component of
\begin{equation}
{\cal H}_{e}^{\mu} = - \frac{1}{2 c^2 \mu_{0}} \, C_{\nu} F^{\nu \mu}.
\label{elmaghel13}
\end{equation}
The divergence of ${\cal H}_{m}^{\mu}$ and ${\cal H}_{e}^{\mu}$ is related to the time conservation of both helicities,
\begin{eqnarray}
\partial_{\mu} {\cal H}_{m}^{\mu} &=& \frac{1}{4 c \mu_{0}} \, F_{\mu \nu} {}^{*}\!F^{\mu \nu} , \nonumber \\ 
\partial_{\mu} {\cal H}_{e}^{\mu} &=& -\frac{1}{4 c \mu_{0}} {}^{*}\!F_{\mu \nu} F^{\mu \nu},
\label{elmaghel14}
\end{eqnarray}
which yields
\begin{eqnarray}
\frac{dh_{m}}{dt} = - \frac{1}{2 c \mu_{0}} \int \left( V \, {\bf B} - {\bf A} \times {\bf E} \right) \cdot d {\bf S} - \frac{1}{c \mu_{0}} \int d^3 r \, {\bf E} \cdot {\bf B} , \nonumber \\  
\frac{dh_{e}}{dt} = - \frac{1}{2 c \mu_{0}} \int \left( V^{\prime} \, {\bf E} + {\bf C} \times {\bf B} \right) \cdot d {\bf S} + \frac{1}{c \mu_{0}} \int d^3 r \, {\bf E} \cdot {\bf B} .
\label{elmaghel15a}
\end{eqnarray}
In the special case that the domain of integration of equations (\ref{elmaghel15a}) is the whole $R^3$ space, and the fields behave at infinity in a way such that the surface integrals in equations (\ref{elmaghel15a}) vanish, we get:
\begin{itemize}
\item If the integral of ${\bf E} \cdot {\bf B}$ is zero, both the
magnetic and the electric helicities are constant during the evolution
of the electromagnetic field.
\item If the integral of ${\bf E} \cdot {\bf B}$ is not zero, the helicities are not constant but they satisfy 
\begin{equation}
\frac{d h_{m}}{dt} = - \frac{d h_{e}}{dt},
\label{elmaghel16}
\end{equation}
so there is an interchange of helicities between the magnetic and electric parts of the field \cite{Arr11}.
\item For every value of the integral of ${\bf E} \cdot {\bf B}$, the electromagnetic helicity $h$, defined as
\begin{equation}
h = h_{m} + h_{e} = \frac{1}{2 c \mu_{0}} \int d^3 r \, {\bf A} \cdot {\bf B} + \frac{\varepsilon_{0}}{2c} \int d^3 r \, 
{\bf C} \cdot {\bf E} , 
\label{elmaghel19}
\end{equation}
is a conserved quantity.
\end{itemize}

If the domain of integration of the equations (\ref{elmaghel15a}) is restricted to a finite volume $\Omega$, then the flux of electromagnetic helicity through the boundary $\partial \Omega$ of the volume is given by
\begin{equation}
\frac{dh}{dt} = - \frac{1}{2 c \mu_{0}} \int_{\partial \Omega} \left[ \left( V \, {\bf B} - {\bf A} \times {\bf E} \right) + \left( V^{\prime} \, {\bf E} + {\bf C} \times {\bf B} \right) \right] \cdot d {\bf S} .
\label{elmaghel20}
\end{equation}
The integrand in the second term of this equation defines a vector density whose components are given by ${\bf S}_{i} = {\cal H}_{m}^{i} + {\cal H}_{m}^{i}$, so that
\begin{equation}
{\bf S} =  \frac{1}{2 c^2 \mu_{0}} \left( V \, {\bf B} - {\bf A} \times {\bf E} + V^{\prime} \, {\bf E} + {\bf C} \times {\bf B} \right) .
\label{elmaghel21}
\end{equation} 
This vector density has been considered as a physically meaningful spin density for the electromagnetic field in vacuum in some references \cite{Bliokh2010,Barnett2010,Barnett2011,Birula2011,Bliokh2013}. In the following, we examine some questions about the relation between the magnetic and electric parts of the helicity and their corresponding magnetic and electric parts of the spin.

\section{Fourier decomposition and helicity basis for the electromagnetic field in vacuum} 
In this section we will write the electromagnetic fields in terms of the helicity basis which will be very useful for obtaining the results and computations presented in the following sections.

The electric and magnetic fields can be decomposed in Fourier terms,
\begin{eqnarray}
{\bf E} ({\bf r}, t) = \frac{1}{(2 \pi)^{3/2}} \int d^3 k \left( {\bf E}_{1} ({\bf 
k}) e^{-ikx} +
{\bf E}_{2} ({\bf k}) e^{ikx)} \right) , 
\nonumber \\
{\bf B} ({\bf r}, t) = \frac{1}{(2 \pi)^{3/2}} \int d^3 k \left( {\bf B}_{1} ({\bf 
k}) e^{-ikx} +
{\bf B}_{2} ({\bf k}) e^{ikx} \right) , 
\label{elmag24}
\end{eqnarray}
where we have introduced the four-dimensional notation $k x = \omega t - {\bf k} \cdot {\bf r}$, with $\omega = k c$.

For the vector potentials we need to fix a gauge. In the Coulomb gauge, the vector potentials are chosen so that $V=0$, $\nabla \cdot {\bf A} = 0$, $V^{\prime} = 0$, $\nabla \cdot {\bf C} = 0$. Then, they satisfy the relations
\begin{eqnarray}
{\bf B} = \nabla \times {\bf A} = \frac{1}{c^2} \, \frac{\partial {\bf C}}{\partial t} , \nonumber \\
{\bf E} = \nabla \times {\bf C} = - \frac{\partial {\bf A}}{\partial t} . \label{elmaghel324}
\end{eqnarray}
One can write for them the following Fourier decomposition,
\begin{eqnarray}
{\bf A} ({\bf r}, t) = \frac{1}{(2 \pi)^{3/2}} \int d^3 k \, \left[ e^{-i k x} \, \bar{{\bf a}} ({\bf k}) + e^{i k x} \,  {\bf a} ({\bf k}) \right] , \nonumber \\
{\bf C} ({\bf r}, t) = \frac{c}{(2 \pi)^{3/2}} \int d^3 k \, \left[ e^{-i k x} \, \bar{{\bf c}} ({\bf k}) + e^{i k x} \,  {\bf c} ({\bf k}) \right] , \label{elmaghel325}
\end{eqnarray}
where the factor $c$ in ${\bf C}$ is taken by dimensional reasons and $\bar{{\bf a}},\bar{{\bf c}}$ denotes the complex conjugate of ${\bf a},{\bf c}$ respectively. Taking time derivatives, and using the Coulomb gauge conditions (\ref{elmaghel324}), 
\begin{eqnarray}
{\bf E} = - \frac{\partial {\bf A}}{\partial t} =  \frac{1}{(2 \pi)^{3/2}} \int d^3 k \, \left[ e^{-i k x} \, (i k c) \, \bar{{\bf a}} ({\bf k}) - e^{i k x} \, (i k c) \, {\bf a} ({\bf k}) \right] , \nonumber \\
{\bf B} = \frac{1}{c^2} \, \frac{\partial {\bf C}}{\partial t} = \frac{1}{(2 \pi)^{3/2}} \int d^3 k \, \left[ - e^{-i k x} \, (i k) \, \bar{{\bf c}} ({\bf k}) + e^{i k x} \, (i k) \, {\bf c} ({\bf k}) \right] . \label{elmaghel24}
\end{eqnarray}
and by comparison with equations (\ref{elmag24}), one can get the values for ${\bf a} ({\bf k})$ and ${\bf c} ({\bf k})$.

The helicity Fourier components appear when the vector potentials ${\bf A}$ and ${\bf C}$, in the Coulomb gauge, are written as a combination of circularly polarized plane waves \cite{Ynd}, as
\begin{eqnarray}
{\bf A} ({\bf r}, t) = \frac{\sqrt{\hbar c \mu_{0}}}{(2 \pi )^{3/2}} \int \frac{d^{3} k}{\sqrt{2 k}} \left[ e^{-i k x} \left( a_{R} ({\bf k})
{\bf e}_{R} ({\bf k}) + a_{L} ({\bf k}) {\bf e}_{L} ({\bf k}) \right)  + C.C \right] , \nonumber \\
{\bf C} ({\bf r}, t) = \frac{c \sqrt{\hbar c \mu_{0}}}{(2 \pi )^{3/2}} \int \frac{d^{3} k}{\sqrt{2 k}} \left[ i \, e^{-i k x} \left( a_{R} 
({\bf k}) {\bf e}_{R} ({\bf k}) - a_{L} ({\bf k}) {\bf e}_{L} ({\bf k}) \right) + C.C\right].
\label{elmaghel34}
\end{eqnarray}
where $\hbar$ is the Planck constant and $C.C$ means complex conjugate. The Fourier components in the helicity basis are given by the unit vectors ${\bf e}_{R} ({\bf k})$, ${\bf e}_{L} ({\bf k})$, ${\bf e}_{k} = {\bf k}/k$, and the helicity components $a_{R} ({\bf k})$, $a_{L} ({\bf k})$ that, in the quantum theory, are interpreted as annihilation operators of photon states with right- and left-handed polarization, respectively. In quantum theory $\bar{a}_{R} ({\bf k})$, $\bar{a}_{L} ({\bf k})$ are creation operators of such states.

In order to simplify the notation, most of the time we will not write explicitly the dependence on ${\bf k}$ of the basis vectors and coefficients, meaning $a_{L} = a_{L} ({\bf k})$, ${\bf e}_{R} = {\bf e}_{R} ({\bf k})$, $a^{\prime}_{L} = a_{L} ({\bf k}^{\prime})$, ${\bf e}^{\prime}_{R} = {\bf e}_{R} ({\bf k}^{\prime})$. 

The unit vectors in the helicity basis are taken to satisfy
\begin{eqnarray}
\bar{{\bf e}}_{R} = {\bf e}_{L} , \, {\bf e}_{R} (- {\bf k}) = - {\bf e}_{L} ({\bf k}) , \, {\bf e}_{L} 
(- {\bf k}) = - {\bf e}_{R} ({\bf k}) , \nonumber \\
{\bf e}_{k} \cdot {\bf e}_{R} = {\bf e}_{k} \cdot {\bf e}_{L} = 0 , \, {\bf e}_{R}
\cdot {\bf e}_{R} = {\bf e}_{L} \cdot {\bf e}_{L} = 0 , \, {\bf e}_{R} \cdot {\bf e}_{L} = 1, \nonumber \\
{\bf e}_{k} \times {\bf e}_{k} = {\bf e}_{R} \times {\bf e}_{R} = {\bf e}_{L}
\times {\bf e}_{L} = 0 , \nonumber \\
{\bf e}_{k} \times {\bf e}_{R} = -i {\bf e}_{R} , \, {\bf e}_{k} \times {\bf e}_{L} = i 
{\bf e}_{L} , \, {\bf e}_{R}\times {\bf e}_{L}= -i {\bf e}_{k}, \label{elmaghel35}
\end{eqnarray}
The relation between the helicity basis and the planar Fourier basis can be obtained by comparing equations (\ref{elmaghel325}) and (\ref{elmaghel34}). Consequently, the electric and magnetic fields of an electromagnetic field in vacuum, and the vector potentials in the Coulomb gauge, can be expressed in this basis as
\begin{eqnarray}
{\bf E} ({\bf r}, t) = \frac{i c \sqrt{\hbar c \mu_{0}}}{(2 \pi )^{3/2}} \int d^{3} k \, \sqrt{\frac{k}{2}} \, \left[ e^{-i k x} \left( a_{R} {\bf e}_{R} + a_{L} {\bf e}_{L} \right)  - e^{i k x} \left( \bar{a}_{R} {\bf e}_{L} + \bar{a}_{L} {\bf e}_{R} \right) \right] \nonumber \\
{\bf B} ({\bf r}, t) = \frac{\sqrt{\hbar c \mu_{0}}}{(2 \pi )^{3/2}} \int d^{3} k \, \sqrt{\frac{k}{2}} \, \left[ e^{-i k x} \left( a_{R} {\bf e}_{R} - a_{L} {\bf e}_{L} \right) + e^{i k x} \left( \bar{a}_{R} {\bf e}_{L} - \bar{a}_{L} {\bf e}_{R} \right) \right] \nonumber \\
{\bf A} ({\bf r}, t) = \frac{\sqrt{\hbar c \mu_{0}}}{(2 \pi )^{3/2}} \int d^{3} k \, \frac{1}{\sqrt{2 k}} \, \left[ e^{-i k x} \left( a_{R} {\bf e}_{R} + a_{L} {\bf e}_{L} \right) + e^{i k x} \left( \bar{a}_{R} {\bf e}_{L} + \bar{a}_{L} {\bf e}_{R} \right) \right] \nonumber \\
{\bf C} ({\bf r}, t) = \frac{i c \sqrt{\hbar c \mu_{0}}}{(2 \pi )^{3/2}} \int d^{3} k \, \frac{1}{\sqrt{2 k}} \, \left[ e^{-i k x} \left( a_{R} {\bf e}_{R} - a_{L} {\bf e}_{L} \right) - e^{i k x} \left( \bar{a}_{R} {\bf e}_{L} - \bar{a}_{L} {\bf e}_{R} \right) \right] \nonumber \\
\label{elmaghel392}
\end{eqnarray}
where the unit vectors satisfy the relations (\ref{elmaghel35}). 

\section{Magnetic and electric helicities in the helicity basis}
In previous section we have introduced the helicity basis and expressed the fields in that basis. In this section we will express the electric and magnetic helicities in the same basis \cite{True96}.

If we use the expressions (\ref{elmaghel392}) the magnetic helicity can be written as 
\begin{eqnarray}
h_{m}&=&\frac{1}{2 c \mu_{0}} \int d^3 r \, {\bf A} \cdot {\bf B} = \frac{\hbar}{4} \int d^3 k \int d^3 k^{\prime} \int \frac{d^3 r}{(2 \pi )^3} \, \sqrt{\frac{k^{\prime}}{k}} \nonumber \\
& &\left[ e^{-i \omega t} e^{i \omega^{\prime} t} e^{i ({\bf k} - {\bf k}^{\prime} ) \cdot {\bf r}} \left( a_{R} {\bf e}_{R} + a_{L} {\bf e}_{L} \right) \cdot \left( \bar{a}^{\prime}_{R} {\bf e}^{\prime}_{L} - \bar{a}^{\prime}_{L} {\bf e}^{\prime}_{R} \right) \right. \nonumber \\
&+&\left. e^{i \omega t} e^{-i \omega^{\prime} t} e^{-i ({\bf k} - {\bf k}^{\prime} ) \cdot {\bf r}} \left( \bar{a}_{R} {\bf e}_{L} + \bar{a}_{L} {\bf e}_{R} \right) \cdot \left( a^{\prime}_{R} {\bf e}^{\prime}_{R} - a^{\prime}_{L} {\bf e}^{\prime}_{L} \right) \right. \nonumber \\
&+&\left. e^{-i \omega t} e^{-i \omega^{\prime} t} e^{i ({\bf k} + {\bf k}^{\prime} ) \cdot {\bf r}} \left( a_{R} {\bf e}_{R} + a_{L} {\bf e}_{L} \right) \cdot \left( a^{\prime}_{R} {\bf e}^{\prime}_{R} - a^{\prime}_{L} {\bf e}^{\prime}_{L} \right) \right.
\nonumber \\
&+&\left. e^{i \omega t} e^{i \omega^{\prime} t} e^{-i ({\bf k} + {\bf k}^{\prime} ) \cdot {\bf r}} \left( \bar{a}_{R} {\bf e}_{L} + \bar{a}_{L} {\bf e}_{R} \right) \cdot \left( \bar{a}^{\prime}_{R} {\bf e}^{\prime}_{L} - \bar{a}^{\prime}_{L} {\bf e}^{\prime}_{R} \right) \right] .
\label{elmaghel40}
\end{eqnarray}
Taking into account the following property of the Dirac-delta function,
\begin{equation}
\int d^3 k^{\prime} \int \frac{d^3 r}{(2 \pi)^{3}} \, e^{-i ({\bf k} - {\bf k}^{\prime}) \cdot {\bf r}} \left( {\bf f} ({\bf k}) \cdot {\bf g} ({\bf k}^{\prime}) \right) = {\bf f} ({\bf k}) \cdot {\bf g} ({\bf k}) , 
\label{elmaghel41}
\end{equation}
and using relations (\ref{elmaghel35}), yields
\begin{eqnarray}
h_{m}&=&\frac{\hbar}{2} \int d^3 k \,
\left( \bar{a}_{R} ({\bf k}) a_{R} ({\bf k}) - \bar{a}_{L} ({\bf k}) a_{L} ({\bf k}) \right) \nonumber \\
&+&\frac{\hbar}{4} \int d^3 k \, e^{-2 i \omega t} \, \left( - a_{R} ({\bf k}) a_{R} (- {\bf k}) + a_{L} ({\bf k}) a_{L} (- {\bf k}) \right) \nonumber \\
&+&\frac{\hbar}{4} \int d^3 k \, e^{2 i \omega t} \, \left( - \bar{a}_{R} ({\bf k}) \bar{a}_{R} (- {\bf k}) + \bar{a}_{L} ({\bf k}) \bar{a}_{L} (- {\bf k}) \right) .
\label{elmaghel42}
\end{eqnarray}
We observe that the magnetic helicity has two contributions: the first term in (\ref{elmaghel42}) is independent of time, and the rest of terms constitute the time-dependent part of the magnetic helicity.

We repeat the same procedure for the electric helicity. The electric helicity can be written as
\begin{eqnarray}
h_{e}&=&\frac{1}{2 c^{3} \mu_{0}} \int d^3 r \, {\bf C} \cdot {\bf E} = \frac{\hbar}{4} \int d^3 k \int d^3 k^{\prime} \int \frac{d^3 r}{(2 \pi )^3} \, \sqrt{\frac{k^{\prime}}{k}} \nonumber \\
& &\left[ e^{-i \omega t} e^{i \omega^{\prime} t} e^{i ({\bf k} - {\bf k}^{\prime} ) \cdot {\bf r}} \left( a_{R} {\bf e}_{R} - a_{L} {\bf e}_{L} \right) \cdot \left( \bar{a}^{\prime}_{R} {\bf e}^{\prime}_{L} + \bar{a}^{\prime}_{L} {\bf e}^{\prime}_{R} \right) \right. \nonumber \\
&+&\left. e^{i \omega t} e^{-i \omega^{\prime} t} e^{-i ({\bf k} - {\bf k}^{\prime} ) \cdot {\bf r}} \left( \bar{a}_{R} {\bf e}_{L} - \bar{a}_{L} {\bf e}_{R} \right) \cdot \left( a^{\prime}_{R} {\bf e}^{\prime}_{R} + a^{\prime}_{L} {\bf e}^{\prime}_{L} \right) \right. \nonumber \\
&-&\left. e^{-i \omega t} e^{-i \omega^{\prime} t} e^{i ({\bf k} + {\bf k}^{\prime} ) \cdot {\bf r}} \left( a_{R} {\bf e}_{R} - a_{L} {\bf e}_{L} \right) \cdot \left( a^{\prime}_{R} {\bf e}^{\prime}_{R} + a^{\prime}_{L} {\bf e}^{\prime}_{L} \right) \right.
\nonumber \\
&-&\left. e^{i \omega t} e^{i \omega^{\prime} t} e^{-i ({\bf k} + {\bf k}^{\prime} ) \cdot {\bf r}} \left( \bar{a}_{R} {\bf e}_{L} - \bar{a}_{L} {\bf e}_{R} \right) \cdot \left( \bar{a}^{\prime}_{R} {\bf e}^{\prime}_{L} + \bar{a}^{\prime}_{L} {\bf e}^{\prime}_{R} \right) \right],
\label{elmaghel43}
\end{eqnarray}
and again using (\ref{elmaghel41}) and ((\ref{elmaghel35}) we get,
\begin{eqnarray}
h_{e}&=&\frac{\hbar}{2} \int d^3 k \,
\left( \bar{a}_{R} ({\bf k}) a_{R} ({\bf k}) - \bar{a}_{L} ({\bf k}) a_{L} ({\bf k}) \right) \nonumber \\
&-&\frac{\hbar}{4} \int d^3 k \, e^{-2 i \omega t} \, \left( - a_{R} ({\bf k}) a_{R} (- {\bf k}) + a_{L} ({\bf k}) a_{L} (- {\bf k}) \right) \nonumber \\
&-&\frac{\hbar}{4} \int d^3 k \, e^{2 i \omega t} \, \left( - \bar{a}_{R} ({\bf k}) \bar{a}_{R} (- {\bf k}) + \bar{a}_{L} ({\bf k}) \bar{a}_{L} (- {\bf k}) \right).
\label{elmaghel44}
\end{eqnarray}

The electromagnetic helicity $h$ in vacuum is the sum of the magnetic and electric helicities. From (\ref{elmaghel42}) and (\ref{elmaghel44}),
\begin{equation}
h = h_{m} + h_{e} = \hbar \int d^3 k \, \left( \bar{a}_{R}({\bf k}) a_{R} ({\bf k}) - \bar{a}_{L} ({\bf k}) a_{L} ({\bf k}) \right).
\label{elmaghel45}
\end{equation}
In Quantum Electrodynamics, the integral in the right hand side of
equation (\ref{elmaghel45}) is interpreted as the helicity operator,
that subtracts the number of left-handed photons from the number of
right-handed photons. From the usual expressions
\begin{eqnarray}
N_{R} &=& \int d^3 k \, \bar{a}_{R}({\bf k}) a_{R} ({\bf k}) , \nonumber \\
N_{L} &=& \int d^3 k \, \bar{a}_{L}({\bf k}) a_{L} ({\bf k}) ,
\label{elmaghel45b}
\end{eqnarray}
we can write (\ref{elmaghel45}) as
\begin{equation}
h = \hbar \left( N_{R}-N_{L} \right) .
\label{elmaghel46}
\end{equation}
Consequently, the electromagnetic helicity (\ref{elmaghel19}) is the
classical limit of the difference between the numbers of right-handed
and left-handed photons \cite{Afana,True96,Ran97}.

However, the difference between the magnetic and electric helicities depends on time in general, since
\begin{eqnarray}
\tilde{h} (t) = h_{m} - h_{e} = \frac{\hbar}{2} \int d^3 k \left[ e^{-2 i \omega t} \left( - a_{R}({\bf k}) a_{R} (-{\bf k}) + a_{L} ({\bf k}) a_{L} (- {\bf k}) \right) \right. \nonumber \\ 
+ \left. e^{2 i \omega t} \left( - \bar{a}_{R}({\bf k}) \bar{a}_{R} (-{\bf k}) + \bar{a}_{L} ({\bf k}) \bar{a}_{L} (- {\bf k}) \right)  \right].
\label{elmaghel47}
\end{eqnarray}
so the electromagnetic field is allowed to exchange electric and magnetic helicity components during its evolution. For an account of this phenomena we refer to \cite{Arr11, Arr17}

\section{Magnetic and electric spin in the helicity basis}
Now in this section we are going to express the magnetic and electric spins components of the total angular momentum in the helicity basis. 

Let us consider the spin vector defined by equation
(\ref{elmaghel21}). It can be written as
\begin{equation}
{\bf s} = {\bf s}_{m} + {\bf s}_{e} ,
\label{elmaghel51}
\end{equation} 
where the magnetic part of the spin is defined from the flux of magnetic helicity,
\begin{equation}
{\bf s}_{m} = \frac{1}{2 c^2 \mu_{0}} \int d^3 r \, \left( V \, {\bf B} - {\bf A} 
\times {\bf E} \right) ,
\label{elmaghel52}
\end{equation} 
and the electric spin comes from the flux of the electric helicity,
\begin{equation}
{\bf s}_{e} = \frac{1}{2 c^2 \mu_{0}} \int d^3 r \, \left( V^{\prime} \, {\bf E} 
+ {\bf C} \times {\bf B} \right) .
\label{elmaghel53}
\end{equation}
Note that the electric spin in equation (\ref{elmaghel53}) can be defined only for 
the case of electromagnetism in vacuum, in the same way as the electric helicity is 
defined only in vacuum.

Using the helicity basis of the previous sections, which where calculated in the Coulomb gauge, the magnetic spin can be written as
\begin{eqnarray}
{\bf s}_{m}&=&\frac{1}{2 c^2 \mu_{0}} \int d^3 r \, {\bf E} \times {\bf A} =
\frac{\hbar}{4} \int d^3 k \int d^3 k^{\prime} \int \frac{d^3 r}{(2 \pi )^3} \, \sqrt{\frac{k^{\prime}}{k}} \nonumber \\
& &\left[ i e^{-i \omega t} e^{i \omega^{\prime} t} e^{i ({\bf k} - {\bf k}^{\prime} ) \cdot {\bf r}} \left( a_{R} {\bf e}_{R} + a_{L} {\bf e}_{L} \right) \times \left( \bar{a}^{\prime}_{R} {\bf e}^{\prime}_{L} + \bar{a}^{\prime}_{L} {\bf e}^{\prime}_{R} \right) \right. \nonumber \\
&-&\left. i e^{i \omega t} e^{-i \omega^{\prime} t} e^{-i ({\bf k} - {\bf k}^{\prime} ) \cdot {\bf r}} \left( \bar{a}_{R} {\bf e}_{L} + \bar{a}_{L} {\bf e}_{R} \right) \times \left( a^{\prime}_{R} {\bf e}^{\prime}_{R} + a^{\prime}_{L} {\bf e}^{\prime}_{L} \right) \right. \nonumber \\
&-&\left. i e^{-i \omega t} e^{-i \omega^{\prime} t} e^{i ({\bf k} + {\bf k}^{\prime} ) \cdot {\bf r}} \left( a_{R} {\bf e}_{R} + a_{L} {\bf e}_{L} \right) \times \left( a^{\prime}_{R} {\bf e}^{\prime}_{R} + a^{\prime}_{L} {\bf e}^{\prime}_{L} \right) \right.
\nonumber \\
&+&\left. i e^{i \omega t} e^{i \omega^{\prime} t} e^{-i ({\bf k} + {\bf k}^{\prime} ) \cdot {\bf r}} \left( \bar{a}_{R} {\bf e}_{L} + \bar{a}_{L} {\bf e}_{R} \right) \times \left( \bar{a}^{\prime}_{R} {\bf e}^{\prime}_{L} + \bar{a}^{\prime}_{L} {\bf e}^{\prime}_{R} \right) \right],
\label{elmaghel54}
\end{eqnarray}
and after the same manipulations as in the previous section, using (\ref{elmaghel41}) and (\ref{elmaghel35}), it turns out
\begin{eqnarray}
{\bf s}_{m}&=&\frac{\hbar}{2} \int d^3 k \,
\left( \bar{a}_{R} ({\bf k}) a_{R} ({\bf k}) - \bar{a}_{L} ({\bf k}) a_{L} ({\bf k}) \right) \, {\bf e}_{k} \nonumber \\
&+&\frac{\hbar}{4} \int d^3 k \, e^{-2 i \omega t} \, \left( a_{R} ({\bf k}) a_{R} (- {\bf k}) - a_{L} ({\bf k}) a_{L} (- {\bf k}) \right) \, {\bf e}_{k} \nonumber \\
&+&\frac{\hbar}{4} \int d^3 k \, e^{2 i \omega t} \, \left( \bar{a}_{R} ({\bf k}) \bar{a}_{R} (- {\bf k}) - \bar{a}_{L} ({\bf k}) \bar{a}_{L} (- {\bf k}) \right) \, {\bf e}_{k} .
\label{elmaghel55}
\end{eqnarray}
As in the case of magnetic helicity (\ref{elmaghel42}), the magnetic spin has two contributions: the first term in (\ref{elmaghel55}) is independent of time, while the rest of terms are, in principle, time-dependent.

In a similar way, the electric spin in the helicity basis is
\begin{eqnarray}
{\bf s}_{e}&=&\frac{1}{2 c^2 \mu_{0}} \int d^3 r \, {\bf C} \times {\bf B} =
\frac{\hbar}{4} \int d^3 k \int d^3 k^{\prime} \int \frac{d^3 r}{(2 \pi )^3} \, \sqrt{\frac{k^{\prime}}{k}} \nonumber \\
& &\left[ i e^{-i \omega t} e^{i \omega^{\prime} t} e^{i ({\bf k} - {\bf k}^{\prime} ) \cdot {\bf r}} \left( a_{R} {\bf e}_{R} - a_{L} {\bf e}_{L} \right) \times \left( \bar{a}^{\prime}_{R} {\bf e}^{\prime}_{L} - \bar{a}^{\prime}_{L} {\bf e}^{\prime}_{R} \right) \right. \nonumber \\
&-&\left. i e^{i \omega t} e^{-i \omega^{\prime} t} e^{-i ({\bf k} - {\bf k}^{\prime} ) \cdot {\bf r}} \left( \bar{a}_{R} {\bf e}_{L} - \bar{a}_{L} {\bf e}_{R} \right) \times \left( a^{\prime}_{R} {\bf e}^{\prime}_{R} - a^{\prime}_{L} {\bf e}^{\prime}_{L} \right) \right. \nonumber \\
&+&\left. i e^{-i \omega t} e^{-i \omega^{\prime} t} e^{i ({\bf k} + {\bf k}^{\prime} ) \cdot {\bf r}} \left( a_{R} {\bf e}_{R} - a_{L} {\bf e}_{L} \right) \times \left( a^{\prime}_{R} {\bf e}^{\prime}_{R} - a^{\prime}_{L} {\bf e}^{\prime}_{L} \right) \right.
\nonumber \\
&-&\left. i e^{i \omega t} e^{i \omega^{\prime} t} e^{-i ({\bf k} + {\bf k}^{\prime} ) \cdot {\bf r}} \left( \bar{a}_{R} {\bf e}_{L} - \bar{a}_{L} {\bf e}_{R} \right) \times \left( \bar{a}^{\prime}_{R} {\bf e}^{\prime}_{L} - \bar{a}^{\prime}_{L} {\bf e}^{\prime}_{R} \right) \right] ,
\label{elmaghel56}
\end{eqnarray}
that after integrating in ${\bf k}^{\prime}$ gives
\begin{eqnarray}
{\bf s}_{e}&=&\frac{\hbar}{2} \int d^3 k \,
\left( \bar{a}_{R} ({\bf k}) a_{R} ({\bf k}) - \bar{a}_{L} ({\bf k}) a_{L} ({\bf k}) \right) \, {\bf e}_{k} \nonumber \\
&+&\frac{\hbar}{4} \int d^3 k \, e^{-2 i \omega t} \, \left( - a_{R} ({\bf k}) a_{R} (- {\bf k}) + a_{L} ({\bf k}) a_{L} (- {\bf k}) \right) \, {\bf e}_{k} \nonumber \\
&+&\frac{\hbar}{4} \int d^3 k \, e^{2 i \omega t} \, \left( - \bar{a}_{R} ({\bf k}) \bar{a}_{R} (- {\bf k}) + \bar{a}_{L} ({\bf k}) \bar{a}_{L} (- {\bf k}) \right) \, {\bf e}_{k} .
\label{elmaghel57}
\end{eqnarray}
Finally, the spin of the electromagnetic field in vacuum is, according to equation (\ref{elmaghel51}),
\begin{equation}
{\bf s} = {\bf s}_{m} + {\bf s}_{e} = \hbar \int d^3 k \,
\left( \bar{a}_{R} ({\bf k}) a_{R} ({\bf k}) - \bar{a}_{L} ({\bf k}) a_{L} ({\bf k}) \right) \, {\bf e}_{k} ,
\label{elmaghel58}
\end{equation}
an expression that is equivalent to the well-known result in Quantum Electrodynamics \cite{Ynd}.

We can compute, as we did for the helicity, the difference between the magnetic and electric parts of the spin,
\begin{eqnarray}
\tilde{{\bf s}} (t) = {\bf s}_{m} - {\bf s}_{e} = \frac{\hbar}{2} \int d^3 k \left[ e^{-2 i \omega t} \left( a_{R}({\bf k}) a_{R} (-{\bf k}) - a_{L} ({\bf k}) a_{L} (- {\bf k}) \right) \right. \nonumber \\ 
+ \left. e^{2 i \omega t} \left( \bar{a}_{R}({\bf k}) \bar{a}_{R} (-{\bf k}) - \bar{a}_{L} ({\bf k}) \bar{a}_{L} (- {\bf k}) \right)  \right] \, {\bf e}_{k} .
\label{elmaghel59}
\end{eqnarray}
Note the similarity in the integrands of the difference between helicities (\ref{elmaghel47}) and the difference between spins (\ref{elmaghel59}). Both have one term proportional to the complex quantity
\begin{equation}
f ({\bf k}) = a_{R}({\bf k}) a_{R} (-{\bf k}) - a_{L} ({\bf k}) a_{L} (- {\bf k}) , 
\label{elmaghel60}
\end{equation}
and another term proportional to the complex conjugate of $f({\bf k})$. It is obvious that $f({\bf k})$ is an even function of the wave vector ${\bf k}$. This means, in particular, that the integral (\ref{elmaghel59}) is identically zero, so the spin difference satisfies
\begin{equation}
\tilde{{\bf s}} (t) = 0.
\label{elmaghel61}
\end{equation}
Thus that we arrive at the following result for any electromagnetic field in vacuum,
\begin{equation}
{\bf s}_{m} = {\bf s}_{e} = \frac{1}{2} \, {\bf s} = \frac{\hbar}{2} \int d^3 k \,
\left( \bar{a}_{R} ({\bf k}) a_{R} ({\bf k}) - \bar{a}_{L} ({\bf k}) a_{L} ({\bf k}) \right) \, {\bf e}_{k} .
\label{elmaghel62}
\end{equation}
This conclusion coincides with the results found in \cite{Barnett2011}.

So while the magnetic and electric spins are equal in electromagnetism in vacuum, in general this fact does not apply for the magnetic and electric helicities, as we have seen in the previous section. These results have been obtained in the framework of standard classical electromagnetism in vacuum but they are also compatible with the suggestion done by Bliokh of a dual theory of electromagnetism \cite{Bliokh2013}.

\section{The angular momentum decomposition for non-null toroidal electromagnetic fields}
In this section we calculate explicitly and analytically the spin-angular decomposition of a whole class of electromagnetic fields in vacuum without using any paraxial approximation.

We will use the knotted non-null torus class\cite{Arr15,Arr17}. These fields are exact solutions of Maxwell equations in vacuum with the property that, at a given time $t = 0$, all pairs of lines of the field ${\bf B} ({\bf r},0)$ are linked torus knots, and that the linking number is the same for all the pairs. Similarly, for the electric field at the initial time ${\bf E} ({\bf r},0)$,  all pairs of lines are linked torus knots and the linking number is the same for all the pairs.

We take a four positive integers tuplet $(n, m, l, s)$. It is possible to find an initial magnetic field such that all its magnetic lines are $(n,m)$ torus knots. The linking number of every two magnetic lines at $t=0$ is equal to $nm$. Also we can find an initial electric field such that all the electric lines are $(l,s)$ torus knots and at $t=0$. At that time, the linking number of the electric field lines is equal to $ls$. We can assure that property at $t=0$, due to the fact that topology may change during time evolution if one of the integers $(n, m, l, s)$ is different to any of the others (for details, we refer the interested reader to \cite{Arr15}). The magnetic and electric helicities also may change if the integer tuplet is not proportional to $(n,n,l,l)$. In these cases, the electromagnetic fields interchange the magnetic and electric helicities during their time evolution. 

We define the dimensionless coordinates $(X, Y, Z, T)$ which are related to the physical ones $(x, y, z, t)$ by $(X, Y, Z, T) = (x, y, z, c t)/L_{0}$, and $r^2 /L_{0}^2 =(x^2 + y^2 +z^2)/L_{0}^2 = X^2 + Y^2 +Z^2 =R^2$. The length scale $L_{0}$ can be chosen to be the mean quadratic radius of the energy distribution of the electromagnetic field. The set of non-null torus electromagnetic knots can be written as  
\begin{eqnarray}
{\bf B} ({\bf r}, t) &=& \frac{\sqrt{a}}{\pi L_{0}^2} \, \frac{Q \, {\bf H}_{1} + P \, {\bf H}_{2}}{(A^2 + T^2 )^3} \label{knot10} \\
{\bf E} ({\bf r}, t) &=& \frac{\sqrt{a} c}{\pi L_{0}^2} \, \frac{Q \, {\bf H}_{4} - P \, {\bf H}_{3}}{(A^2 + T^2 )^3} \label{knot11}
\end{eqnarray}
where $a$ is a constant related to the energy of the electromagnetic field,
\begin{equation}
A = \frac{1+R^2 -T^2}{2} , \, \, \, P = T (T^2 - 3 A^2) , \, \, \, Q = A (A^2 - 3 T^2),
\label{knot12}
\end{equation}
and
\begin{eqnarray}
{\bf H}_{1} = \left( -n \, XZ + m \, Y + s \, T \right) \, {\bf u}_{x} + \left( -n \, YZ -m \, X -l \, TZ \right) \, {\bf u}_{y} \nonumber \\
+ \left( n \, \frac{-1 - Z^2 + X^2 + Y^2 +T^2}{2} + l \, TY \right) \, {\bf u}_{z} . \label{knot13} \\
{\bf H}_{2} = \left( s \, \frac{1+X^2 -Y^2-Z^2-T^2}{2} -m \, TY \right) \, {\bf u}_{x} \nonumber \\
+ \left( s \, XY - l \, Z + m \, TX \right) \, {\bf u}_{y} + \left( s \, XZ + l \, Y + n \, T \right) \, {\bf u}_{z} . \label{knot14} \\
{\bf H}_{3} = \left( -m \, XZ + n \, Y + l \, T \right) \, {\bf u}_{x} + \left( -m \, YZ -n \, X -s \, TZ \right) \, {\bf u}_{y} \nonumber \\ + \left( m \, \frac{-1 - Z^2 + X^2 + Y^2 +T^2}{2} + s \, TY \right) \, {\bf u}_{z} . \label{knot15} \\
{\bf H}_{4} = \left( l \, \frac{1+X^2 -Y^2-Z^2-T^2}{2} - n \, TY \right) \, {\bf u}_{x} \nonumber \\ 
+ \left( l \, XY - s \, Z + n \, TX \right) \, {\bf u}_{y} + \left( l \, XZ + s \, Y + m \, T \right) \, {\bf u}_{z}. \label{knot16}
\end{eqnarray}
The energy ${\cal E}$, linear momentum ${\bf p}$ and total angular momentum ${\bf J}$ of these fields are
\begin{eqnarray}
{\cal E} &=& \int \left( \frac{\varepsilon_{0} \, E^2}{2} + \frac{B^2}{2 \mu_{0}} \right) d^3r= \frac{a}{2 \mu_{0} L_{0}} (n^2 + m^2 + l^2 + s^2) \label{knot19} \\
{\bf p} &=& \int \varepsilon_{0} \, {\bf E} \times {\bf B}\, d^3r = \frac{a}{2 c \mu_{0} L_{0}} \, (l n + m s) \, {\bf u}_{y} \label{knot20} \\
{\bf J} &=& \int \varepsilon_{0} \, {\bf r} \times \left(  {\bf E} \times {\bf B} \right)d^3r = \frac{a}{2 c \mu_{0}} \, (l m + n s) \, {\bf u}_{y} \label{knot20b} 
\end{eqnarray}

To study the interchange between the magnetic and electric helicities and the spins, we first need the Fourier transforms of the fields in the helicity basis. Following the prescription given in Section 3, we get
\begin{eqnarray}
a_{R} {\bf e}_{R} + a_{L} {\bf e}_{L} &=& \sqrt{\frac{a}{\hbar c \mu_{0}}} \, \frac{L_{0}^{3/2}}{2 \sqrt{\pi}} \, \frac{e^{-K}}{\sqrt{K}} \, \times \nonumber \\
& &\left[ \frac{m}{K} \left( K_{x} K_{z}, K_{y} K_{z}, - K_{x}^2 -K_{y}^2 \right) + s \left( 0, K_{z}, - K_{y} \right) \right] \nonumber \\
&+& i \, \left[ \frac{l}{K} \left( -K_{y}^2 - K_{z}^2, K_{x}K_{y},K_{x}K_{z} \right) + n \left( -K_{y}, K_{x},0 \right) \right] \label{knot21} \\
a_{R} {\bf e}_{R} - a_{L} {\bf e}_{L} &=& \sqrt{\frac{a}{\hbar c \mu_{0}}} \, \frac{L_{0}^{3/2}}{2 \sqrt{\pi}} \, \frac{e^{-K}}{\sqrt{K}} \, \times \nonumber \\
& &\left[ \frac{n}{K} \left( K_{x} K_{z}, K_{y}K_{z}, -K_{x}^2 - K_{y}^2 \right) + l \left( 0, K_{z}, -K_{y} \right) \right] \nonumber \\
&+& i \, \left[ \frac{s}{K} \left( - K_{y}^2 -K_{z}^2, K_{x} K_{y}, K_{x} K_{z} \right) + m \left( -K_{y}, K_{x}, 0 \right) \right] 
\label{knot22} 
\end{eqnarray}
In these expressions, we have introduced the dimensionless Fourier space coordinates $(K_{x},K_{y},K_{z})$, related to the dimensional Fourier space coordinates $(k_{x}, k_{y},k_{z})$ according to
\begin{equation}
(K_{x},K_{y},K_{z}) = L_{0} (k_{x},k_{y},k_{z}) , \,\,\, K = L_{0} k = \frac{L_{0} \omega}{c}.
\label{knot23}
\end{equation}
The electromagnetic helicity (\ref{elmaghel45}) of the set of non-null torus electromagnetic knots results
\begin{equation}
h = \hbar \int d^3 k \, \left( \bar{a}_{R}({\bf k}) a_{R} ({\bf k}) - \bar{a}_{L} ({\bf k}) a_{L} ({\bf k}) \right) = \frac{a}{2 c \mu_{0}} ( n m + l s ) ,
\label{knot24}
\end{equation}
and the difference between the magnetic and electric helicities is
\begin{eqnarray}
\tilde{h} (t) &=& h_{m} - h_{e} = \frac{\hbar}{2} \int d^3 k \left[ e^{-2 i \omega t} \left( - a_{R}({\bf k}) a_{R} (-{\bf k}) + a_{L} ({\bf k}) a_{L} (- {\bf k}) \right) \right. \nonumber \\ 
              &+& \left. e^{2 i \omega t} \left( - \bar{a}_{R}({\bf k}) \bar{a}_{R} (-{\bf k}) + \bar{a}_{L} ({\bf k}) \bar{a}_{L} (- {\bf k}) \right)  \right]\nonumber\\
&=& \frac{a}{2 c \mu_{0}} (n m - l s) \frac{1- 6 T^2 + T^4}{(1 + T^2)^4},
\label{knot25}
\end{eqnarray}
where we recall that $T = c t/L_{0}$. Results (\ref{knot24}) and (\ref{knot25}) coincide with the computations done in \cite{Arr15} using different procedures.

Now consider the spin in equation (\ref{elmaghel58}). For the set of non-null torus electromagnetic knots, we get 
\begin{equation}
{\bf s} = \hbar \int d^3 k \,
\left( \bar{a}_{R} ({\bf k}) a_{R} ({\bf k}) - \bar{a}_{L} ({\bf k}) a_{L} ({\bf k}) \right) \, {\bf e}_{k} = \frac{a}{4 c \mu_{0}} (m l + n s)\, {\bf u}_{y}.
\label{knot26}
\end{equation}
Notice that this value of spin is equal to one half of the value of the total angular momentum obtained in equation (\ref{knot20b}). Thus the orbital angular momentum of this set of electromagnetic fields has the same value as the spin angular momentum,
\begin{equation}
{\bf L} = {\bf s} = \frac{1}{2} \, {\bf J} .
\label{knot27}
\end{equation}
The difference between the magnetic and the electric spin can also be computed through equation (\ref{elmaghel59}). The results is
\begin{eqnarray}
\tilde{{\bf s}} (t) = {\bf s}_{m} - {\bf s}_{e} = \frac{\hbar}{2} \int d^3 k \left[ e^{-2 i \omega t} \left( a_{R}({\bf k}) a_{R} (-{\bf k}) - a_{L} ({\bf k}) a_{L} (- {\bf k}) \right) \right. \nonumber \\ 
+ \left. e^{2 i \omega t} \left( \bar{a}_{R}({\bf k}) \bar{a}_{R} (-{\bf k}) - \bar{a}_{L} ({\bf k}) \bar{a}_{L} (- {\bf k}) \right)  \right] \, {\bf e}_{k} = {\bf 0}.
\label{knot28}
\end{eqnarray}
In consequence, even if the magnetic and electric helicities depend on time for this set of electromagnetic fields, the magnetic and electric parts of the spin are time independent, satisfying the results found in equation (\ref{elmaghel62}) for general electromagnetic fields in vacuum. Both are equal, and satisfy
\begin{equation}
{\bf s}_{m} = {\bf s}_{e} = \frac{1}{2} \, {\bf s} = \frac{a}{8 c \mu_{0}} (m l + n s) \,{\bf u}_{y}.
\label{knot29}
\end{equation}

\section{Same spin-orbital decomposition with different behavior in the helicities}
In this section we will present two particular cases examples where the spin and orbital decomposition of the angular momentum are equal in both cases, while the helicities are constant and non-constant respectively. We will see that the angular momentum density evolves differently in each case.

\begin{figure}
  \centering
  \includegraphics[width=0.3\linewidth]{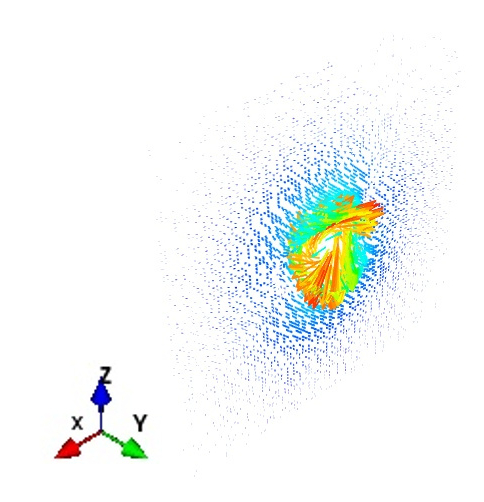}
  \includegraphics[width=0.3\linewidth]{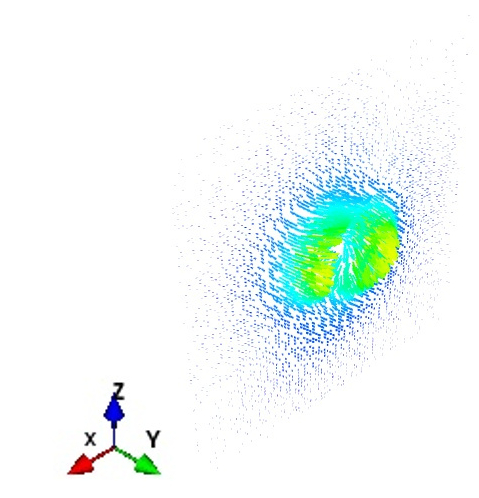}
  \includegraphics[width=0.3\linewidth]{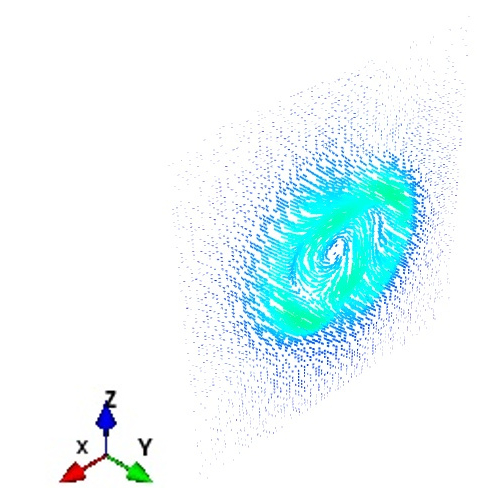}\\
  \includegraphics[width=0.3\linewidth]{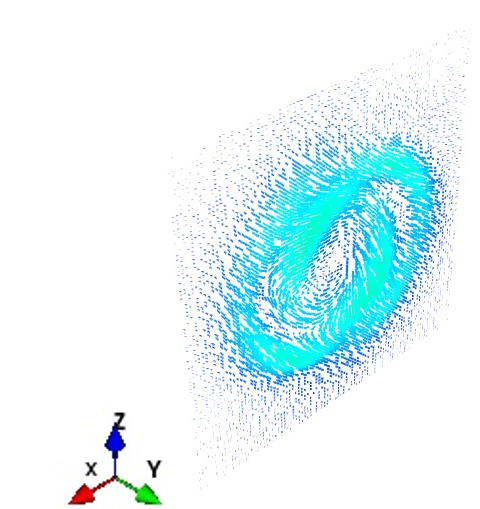}
  \includegraphics[width=0.3\linewidth]{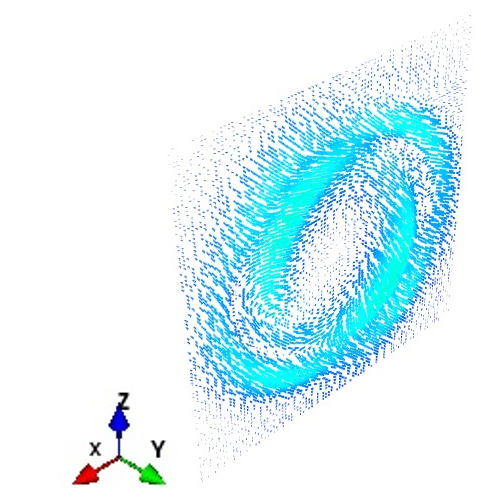}
  \includegraphics[width=0.3\linewidth]{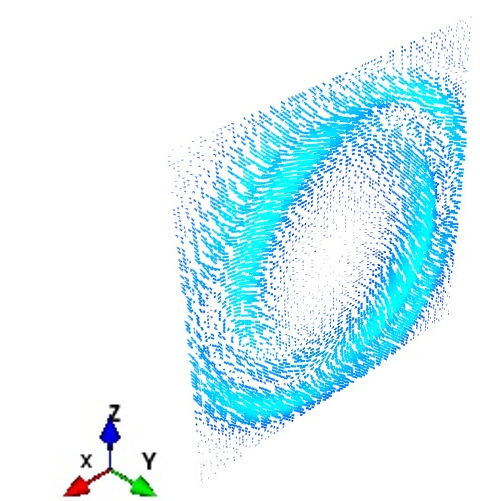}
  \caption{The angular momentum density ${\bf j}$ at times $T =0, 0.5, 1, 1.5, 2, 2.5$, for the electromagnetic field given by the set $(n,m,l,s) = (5,3,5,3)$. The vector field is sampled at the plane $XZ$. In the case depicted in the figure, the magnetic helicity is equal to the electric helicity and constant in time}
  \label{fig:1}
\end{figure}

In the first case, we take the set $(n,m,l,s) = (5,3,5,3)$ in expressions (\ref{knot10}) and (\ref{knot11}). Thus using (\ref{knot20b}) the total angular momentum is \[{\bf J}=\frac{15a}{ \mu_{0}} \, {\bf u}_{y},\] while the angular density changes in time. In order to visualize the evolution of the angular momentum density, which is given by ${\bf j} ={\bf r} \times \left(  {\bf E} \times {\bf B} \right)$, we plot at different times the vector field sample at the plane $XZ$, as it is depicted in Fig.~\ref{fig:1}.

For this case, the spin-orbital split, as shown in previous section, using (\ref{knot27}), turns out to be
\begin{equation}
  \label{eq:ex1}
  {\bf L}={\bf s}= \frac{15a}{2 \mu_{0}} \, {\bf u}_{y}.
\end{equation}
which remain constant during the time evolution of the field. The magnetic and electric helicities remains also constant, and there is not exchange between them. 

\begin{figure}
  \centering
  \includegraphics[width=0.3\linewidth]{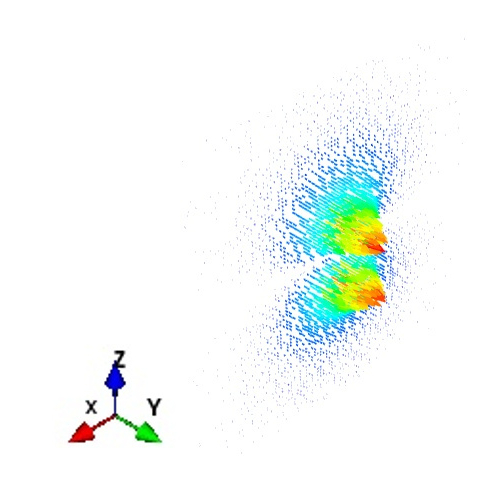}
  \includegraphics[width=0.3\linewidth]{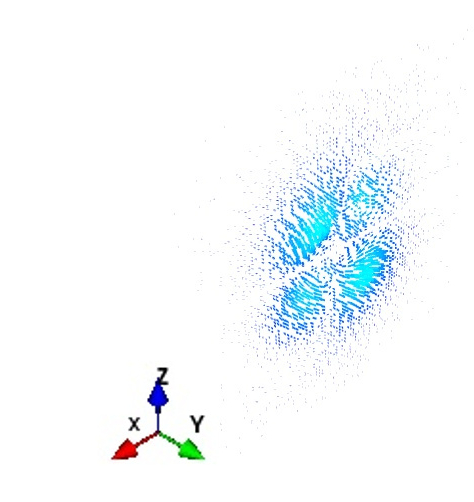}
  \includegraphics[width=0.3\linewidth]{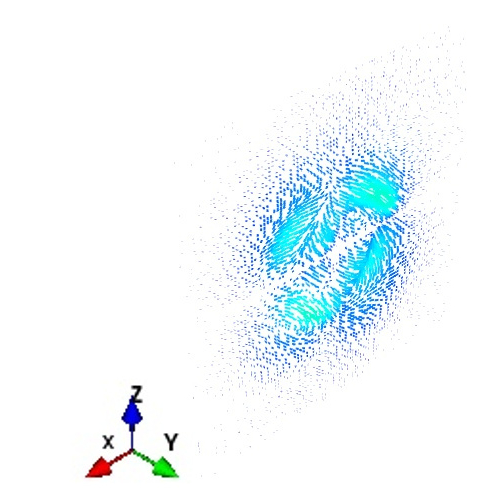}\\
  \includegraphics[width=0.3\linewidth]{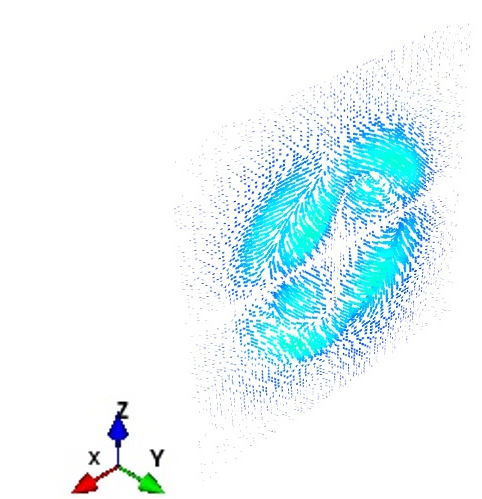}
  \includegraphics[width=0.3\linewidth]{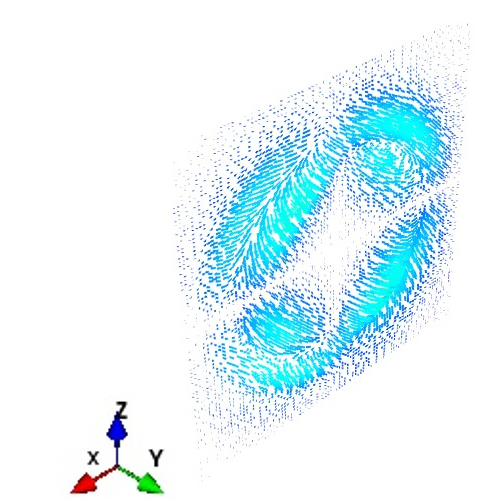}
  \includegraphics[width=0.3\linewidth]{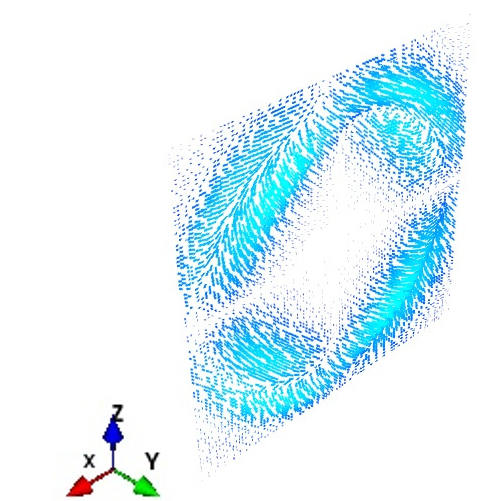}
  \caption{The angular momentum density ${\bf j}$ at times $T=0, 0.5, 1, 1.5, 2, 2.5$, for the electromagnetic field given by the set $(n,m,l,s) = (15,5,0,2)$. The vector field is sampled at the plane $XZ$. In this example, the magnetic and electric helicities are initially different and their values change with time}
  \label{fig:2}
\end{figure}

Now let us take the set $(n,m,l,s) = (15,5,0,2)$ in expressions (\ref{knot10}) and (\ref{knot11}). The electromagnetic field obtained with this set of integers has the same value of the total angular momentum that the previous case, and the same spin-orbital split. However, in this case 
the magnetic and electric helicities are time-dependent, satisfying equation (\ref{knot25}). The time evolution of the angular momentum density is different from the case of constant helicities, as we can see in Fig.~\ref{fig:2}. As we did before, we have plotted the field at the plane $XZ$ at same time steps than in the first example. 

In the first example of non-null torus electromagnetic field, the helicities remain constant in time. In the second example, the magnetic helicity is initially different from the electric helicity and both change with time. Even if the spin, orbital and total angular momenta are equal in both examples, we can see in Fig.~\ref{fig:1} and Fig.~\ref{fig:2} that the structure of the total angular momentum density is different. We can speculate that a macroscopic particle which can interact with the angular momentum of the field would behave in the same way in both cases, but if we have a microscopic test particle able to interact with the density of the angular momentum would behave differently.   

\section{Conclusions}
We have calculated analytically and exactly the spin-orbital
decomposition of the angular momentum of a class of electromagnetic
fields beyond the paraxial approximation. A spin density that is dual
in its magnetic and electric contributions has been considered. This
spin density has the meaning of flux of electromagnetic helicity. By
using a Fourier decomposition of the electromagnetic field in vacuum
in terms of circularly polarized waves, called the helicity basis, we
have given explicit expressions for the magnetic and electric
contributions to the spin angular momentum. We have obtained the
results that the magnetic and electrical components of spin remain
constant during the time evolution of the fields. We also have made
use of the helicity basis to calculate the magnetic and electric
helicities.

We have obtained the exact split of the angular momentum into a spin
and orbital components for electromagnetic fields which belong to the
non-null toroidal knotted class \cite{Arr15}. One of main characteristics of that
class is that contains certain degree of linkage of electric and
magnetic lines and can have exchange between the magnetic and
electrical components of the helicity \cite{Arr11}.

We have considered two examples of these non-null knotted electromagnetic fields having the properties that they have the same angular momentum and the same split. They have the same constant values for the orbital and spin components of the angular momentum, the first with constant and equal helicities and the second with time-evolving helicities. The behaviour of the total angular momentum density seems to be different in these two cases. 

In our opinion, the study of this kind of examples with nontrivial helicities may provide a clarification of the role of helicities in the behaviour of angular momentum densities of electromagnetic fields in vacuum. 

\section*{Acknowledgements}
We acknowledge Wolfgang L{\"o}ffler for valuable discussions at the initial stage of this work. 
This work was supported by research grants from the Spanish Ministry of Economy and Competitiveness (MINECO/FEDER) ESP2015-69909-C5-4-R.

\end{document}